\documentclass[12pt,preprint]{aastex}

\shorttitle{Iron lines with reconnection-heated corona}

\received{2004 November 16}
\begin{document}

\title{Iron fluorescent line emission from black hole accretion disks with magnetic reconnection-heated corona}
\author{N. Kawanaka\altaffilmark{1}, S. Mineshige\altaffilmark{1} and K. Iwasawa\altaffilmark{2}}
\altaffiltext{1}{Yukawa Institute for Theoretical Physics, Kyoto University,
Kyoto 606-8502, Japan}
\email{norita@yukawa.kyoto-u.ac.jp}

\altaffiltext{2}{Institute of Astronomy, Madingley Road, Cambridge CB3 0HA, UK}

\begin{abstract}
  We investigate the iron K$\alpha$ fluorescent line produced by hard X-ray photons from magnetic reconnection-heated
 corona.  The hot corona with temperature being about $10^9$K can irradiate the underlying disk with a continuum X-ray
 spectrum produced via thermal Comptonization.  Then the iron atoms in the disk photoelectrically absorb X-ray photons
 and radiate K$\alpha$ line photons.  Therefore, the activity of corona is responsible to the iron line emission
 from the underlying disk.  In previous studies, oversimplified X-ray photon sources are often assumed above the disk in order
 to compute the iron line profile or power-law line emissivity profiles are assumed with an index being a free parameter.

 We adopt the more realistic corona model constructed by Liu et al. in which the corona is heated by magnetic
 energy released through the reconnection of magnetic flux loops and which has no free parameter.  Then the accretion energy is
 dominantly dissipated in the corona, in which X-ray photons are efficiently produced and irradiate the underlying disk.
  We find the local emmisivity of iron line on the disk is approximated as $F_{{\rm K}\alpha}(r)\propto r^{-5}$.  The iron line profiles
 derived from this model give excellent fits to the observational data of MCG-6-30-15 with the profiles derived
 theoretically for $i\sim 30^{\circ }$ for energy band 4-7keV.  Possible origins of line variability are briefly discussed.
\end{abstract}

\section{INTRODUCTION}
Iron K$\alpha$ line is one of the most useful probes of the vicinity of the black hole.  It may be produced by X-ray
 irradiation on cold iron atoms in an optically thick accretion disk.  Although the line width is intrinsically small,
 the line profile is broadened by Doppler effects, and distorted by relativistic beaming and gravitational redshifts,
since the material is orbiting at high velocity in a strong gravitational field.  So it is expected
 that we can get some information about the gravitational field and the accretion flow close to the central black hole
 from the shape of iron line profile (for comprehensive reviews, see Fabian et al. 2000; Reynolds \& Nowak 2003). 

In many previous studies, the relativistic effects that distort the iron line profile have been computed in detail.
  Most of them, however, used very simplified models of iron line emissivity distribution.  Some authors assumed
 a point source above the disk whose position and luminosity vary with time (Ruszkowski 2000; Lu \& Yu 2001), and
 some authors adopted a power-law emissivity distribution on the disk (Fabian et al. 1989; Laor 1991; Kojima 1991; Dov$\check {\rm c}$iak, Karas, \& Yaqoob 2004).
  Although one can reconstruct the observed iron line profiles and
 its time variability with these models, little can be known regarding the fundamental physics which describes the origins
 of X-ray irradiation in the black hole accretion disk system.  In order to get information about the physical processes
 around the black hole by analyzing the observed iron line profile, such overidealized models are of limited use,
 so one should construct more realistic X-ray source model taking into account the physics of radiation and magnetic processes.

So what is the primal X-ray source and how does it illuminate the disk? 
If the mass accretion rate is sufficiently high, the disk-corona structure may be established.
  In this case, one can expect that the X-ray photons which impinge the cold disk would be produced in the corona through inverse-Compton
 scattering of thermal soft photons from the disk.  In fact, X-ray spectral features observed in ordinary active galactic nuclei (AGNs) and some galactic
 black hole candidates (GBHCs) show  that there is hot gas coexisting with cold matter in the vicinity of an accreting black hole (
see Mushotzky, Done \& Pounds 1993) .  There are many studies about disk-corona system (e.g., Haardt \& Maraschi 1991, 1993).
However, little attempts to compute the line profile with such corona as a realistic X-ray source has been done because the
 coronal heating process remains unclear.

In this study,  we adopt the corona model in which the magnetic flux loop emerging from the disk reconnects with other loops and
heat the corona to a temperature around $10^9$K (Liu, Mineshige, \& Shibata 2002, hereafter LMS02) .  It is known that emergent spectra calculated
 from this model through the Monte Carlo simulations are close to the observed spectra in Seyfert galaxies and radio-quiet
 QSOs (Liu, Mineshige, \& Ohsuga 2003, hereafter LMO03).  The advantage of this model is that one can compute the iron line profile
 without any adjustment of free parameter but a black hole mass $M$ and an accretion rate $\dot{M}$. We assume that the iron
 line emissivity distribution on the disk is determined by the X-ray continuum emission in the corona right above the point of interest.

 The plan of this paper is as follows: in \S 2, we describe the magnetic reconnection-heated corona model which is used in our study.
  Then we calculate the iron line profile predicted from this corona model and make some fits with observations in \S 3.  Discussion and
 conclusions are given in \S 4. 

\section{THE DISK-CORONA MODEL}
We consider the disk-corona model constructed by LMS02, in which the corona is assumed to be plain-parallel and to be
 coupled tightly with underlying Shakura \& Sunyaev (1973) disk (with zero-torque inner boundary condition).  Magnetic flux loops generated in the disk emerge into the corona by the magnetic 
buoyancy, and reconnect with other loops.  As a result, the magnetic energy in the loops is released in the corona as thermal
 energy.  This coronal heat is cooled down by Compton scattering and thus the energy balance in the magnetic flux tube is attained.
 
\begin{eqnarray}
\frac{B^2}{4\pi}V_A\approx \frac{4k_{\rm B}T}{m_e c^2} n \sigma_{\rm T} c U_{\rm rad} \lambda_{\tau} l, \label{comptoncool}
\end{eqnarray}
where $V_A$ is the Alfv\'{e}n speed, $U_{\rm rad}=U_{\rm rad}^{\rm in}+U_{\rm rad}^{\rm re}$ is the soft photon
 field to be Compton scattered from both the intrinsic disk and reprocessed radiation, $l$ is the length of the
 magnetic loop. Here, $\lambda_{\tau}$ is introduced in order to take into account the isotropy of incident photons.  If
 $\lambda_{\tau}=1$, the effective optical depth in the corona is equal to the vertical scattering optical depth
 $\tau=n \sigma_T l$.  In a plane parallel geometry, however, the actual optical depth should be larger than the vertical
 one since the incident photons would not be necessary injected in vertical direction and would undergo a longer path than $l$.
  Then if $\lambda_{\tau}$ were set to be unity, we would overestimate the coronal temperature, and therefore too large
 upscattered luminosity would be derived by Monte Carlo simulation (see below).  In LMO03 and our study, the value of
 $\lambda_{\tau}$ is determined so that the integrated upward luminosity is equal to the released gravitational
 energy(see LMO03); other parameters and constants have their standard meanings.

 If the density of corona is not high enough for Compton cooling, heat is conducted by electrons from the corona to the
 chromosphere, which is dominantly cooled down by evaporation of plasma in the disk:

\begin{eqnarray}
\frac{k_0 T^{7/2}}{l}\approx \frac{\gamma }{\gamma -1}nkT \left( \frac{kT}{m_{\rm H}} \right) ^{1/2},
\end{eqnarray}
where $k_0\approx 10^{-6} {\rm ergs~cm^{-1}~s^{-1}~K^{-7/2}} and \gamma=5/3$.

Moreover, we assume equipartition of gas energy and magnetic energy in the disk,

\begin{eqnarray}
\beta \equiv \frac{n_{\rm disk} k T_{\rm disk}}{B^2/8\pi}\sim 1
\end{eqnarray}
(although the results are not sensitive to $\beta$-values).

With these assumptions, we can derive the equations which
 concern the fraction of accretion energy dissipated in the corona $f$, which has been treated as a fitting parameter in previous
 disk corona models.  For a gas pressure-dominant disk, the soft photon field is dominated by the reprocessed coronal radiations,
 
\begin{eqnarray}
U_{\rm rad}\approx 0.4\lambda _{u} U_{B},
\end{eqnarray}
where $U_{B}$ is the magnetic field energy density.  This equation requires some explanation.  Haardt \& Maraschi (1991)
 showed that the fraction of Compton-scattered photons which
 illuminate the underlying disk is about 0.5-0.6, and the albedo of the disk is about 0.1-0.2 for a specific range of parameters.
  Then, in LMS02, the reprocessed soft photon energy density is set to be $U_{\rm rad}\approx 0.4U_{B}$.  However, the precise
 value of $U_{\rm rad}$ generally depend on the coronal temperature and density.  We introduce the parameter $\lambda_u$ in order to
 take into account this effect and adopt the value so the downward luminosity is equal to the luminosity of soft photons reprocessed
 in the cold disk (see LMO03).

 Then we get an equation,

\begin{eqnarray}
f&=&4.70 \times 10^4 (1-f)^{11/10} \alpha _{0.1}^{-99/80} \beta _1^{-11/8} \lambda_{\tau}^{1/4} \lambda_u^{1/4} m_8^{11/80}\nonumber \\
&&\times \left( \dot {m}_{0.1} \phi \right) ^{1/10} r_{10}^{-81/160} l_{10}^{3/8}, \label {gas}
\end{eqnarray}
where $\phi \equiv 1-(R_{\ast }/R)^{1/2}$ and $R_{\ast}$ is taken to be the last stable orbit $3R_{\rm S}$; and $\alpha _{-1}$, $\beta$, $m_8$,
 $\dot {m}_{-1}$, $r_{10}$, $l_{10}$ are the viscous coefficient, the plasma beta, the black hole mass, the accretion rate, the distance,
 the length of the magnetic loop in units of $0.1$, $1$, $10^8 M_{\odot }$, $0.1 \dot {M}_{\rm {Edd}}$, $10R_{\rm S}$, $10R_{\rm S}$, respectively.

By solving equation (\ref {gas}) for $f$ for a given black hole mass and accretion rate, we can calculate the coronal
 quantities (as well as the disk quantities) at any distances.

Figure 1 shows the coronal structures along distance for $M=10^8 M_{\odot }$ and $\dot{M}=0.1\dot{M}_{\rm Edd}$.  The coronal temperature is 
$\sim 10^9 {\rm K}$ and the density is $\sim 10^9 {\rm cm}^{-3}$.  In such corona, continuum X-ray photons are efficiently
 produced via inverse Comptonization, and part of them are upscattered and escape from the disk-corona system.
  The emergent spectrum
 produced by such photons were calculated from Monte Carlo simulations in LMO03 (see also \S 3).  They showed that the X-ray spectral
 indices of the calculated
spectrum between 2 and 20 keV are around 1.1, which are close to that of the observed spectra of Seyfert galaxies and QSOs.  

On the other hand, there are also some photons which are backscattered in the corona and do not escape the disk-corona
 system. They impinge the underlying disk and drive iron line fluorescence in it.  Given such a condition, we can calculate
 the line profile from this disk-corona system.  We will show the computational method and its results in detail in the
 next section.

\section{COMPUTATIONS AND RESULTS}
\subsection{iron line emissivity from disk} 
  For deriving the iron fluorescence emission law on the disk, it is necessary to derive the X-ray spectrum constructed by the photons
 downscattered in the corona.  Since the coronal properties were determined by LMS02 and LMO03, we can calculate the illumination
 spectrum on each radial grid of the disk by Monte Carlo simulations.

  For typical AGN system with a black hole mass
 of $10^8 M_{\odot }$ and an accretion rate of $0.1M_{\rm Edd}$, the magnetic field is $B\sim 10^3{\rm G}$ under
 the assumption of energy equipartition with gas in the disk.  Then we can estimate the energy flux from the corona onto the
 disk using the expression of Eq. (\ref{comptoncool}), and therefore the ionization parameter $\xi $ of the disk;

\begin{eqnarray}
\xi &\equiv &\frac{4\pi F_{X}(r)}{n_{\rm disk}(r)} \nonumber \\
&\leq &\frac{4\pi F(r)}{n_{\rm disk}(r)} \nonumber \\
&\sim  &\frac{4\pi}{n_{\rm disk}(r)}\times \frac{1}{2}\frac{B^2 V_A}{4\pi} \nonumber \\
&\sim &0.1~{\rm ergs~cm~s^{-1}} \ll 100~{\rm ergs~cm~s^{-1}},
\end{eqnarray}
where $F_X(r)$ and $F(r)$ are the X-ray flux and the bolometric flux striking the unit area of the disk at radius $r$,
 respectively. 
Then, according to the investigation of Matt et al. (1993, 1996), we can assume that the disk material is sufficiently
 cold and dense so that the ionization of metals in the disk can be neglected,
 though this is not always the case for the disk which is strongly illuminated by coronal X-ray.

  According to the approximate expression derived by George \& Fabian
 (1991), the number flux of iron fluorescent photons per unit time that emerge from the disk is given by

\begin{eqnarray}
F_{{\rm K}\alpha}=\int_0^{\pi /2} \int_{E_{\rm min}}^{E_{\rm max}} G(E, \theta_{\rm in})N(E, \theta_{\rm in}) dE d\theta_{\rm in}, \label{flux}
\end{eqnarray}
where $\theta_{\rm in}$ is the incident angle, $N(E, \theta_{\rm in})$ is the number spectrum of the incident photons as the function of
 $\theta_{\rm in}$, and
\begin{eqnarray}
G(E, \theta_{\rm in})&=&g(\theta_{\rm in})f(E), \\
g(\theta_{\rm in})&=&10^{-2} \times (6.5-5.6 \cos \theta_{\rm in} +2.2 \cos ^2 \theta_{\rm in}), \\
f(E)&=&7.4\times 10^{-2}+2.5\times \exp \left( -\frac{E-1.8}{5.7} \right)
\end{eqnarray}
(note that this expression is justified only when one adopts the cosmic element abundances).

In equation (\ref{flux}), $E_{\rm min}$ and $E_{\rm max}$ should be equal to 7.1keV, which is the energy of iron K edge, and 30keV, which is
 the upper application limit of George \& Fabian approximation, respectively.  In this way we can derive the fluorescent line
 emissivity on the disk as the function of radius. 

Figure 2 shows the radial dependence of the iron line photon flux on the disk.
  This profile can be fitted to a power-law $\propto r^{-\beta}$ with $\beta \sim 4-5$ down to $r/r_g\approx 6$.  We also take into
 consideration the anisotropy of the intensity of emerging iron line photons with the formula

\begin{eqnarray}
I_{{\rm K}\alpha}(\theta_{\rm out})d\Omega \propto \frac{\cos \theta_{\rm out}}{\pi} \ln \left( 1+\frac{1}{\cos \theta_{\rm out}} \right) d\Omega
\end{eqnarray}
(Basko 1978; Haardt 1993; Ghisellini, Haardt, \& Matt 1994), where $\theta_{\rm out}$ is the angle between the momentum vector of a photon and
 the vector normal to the disk surface in the corotating frame and it should be determined by

\begin{eqnarray}
\cos \theta_{\rm out}=-\frac{p_{\mu} n^{\mu}}{p_{\nu} u^{\nu}},
\end{eqnarray}
where $p_{\mu}, n_{\mu}, u_{\nu}$ are 4-vectors representing the photon's momentum, the surface normal of the disk, the velocity of the disk
 material, respectively. 

\subsection{line profiles}
To calculate the iron line profile from a given line emissivity law, we use the ray-tracing method (Luminet 1979) and take
 into account general relativistic energy shift in calculating the line intensity.  In our model, the parameters
 we can vary are the black hole mass, the mass accretion rate, and the inclination angle.  In fact, however, the calculated line
 profiles are not so sensitive to the first two parameters.  With this reason, we show the line profiles for various viewing angles
 for fixed $M$ and $\dot {M}$.

Figure 3 shows the line profiles observed from different viewing angles.  The line photons are assumed to be emitted from the
 region extending between $6r_g$ and $50r_g$ from the black hole.  With large inclination, the red wing of line profile
 is more enhanced than that calculated with a power-law emissivity $(\propto r^{-3})$ which is the same distribution as
 the energy flux from the standard disk, because the emissivity profile obtained from this corona model is as steep as
 $\propto r^{-5}$ and a large number of iron line photons are emitted from the innermost region where the gravitational
 redshift is significant.

The best-studied observation of iron line emission from an accretion disk is the one by Tanaka et al. (1995) who showed the
 time-averaged iron line profile from the Seyfert 1 galaxy MCG-6-30-15.  It is claimed to be the evidence that there
 exists the strong gravitational field at the center of the galaxy.  Figure 4 displays the best-fit profile
 of MCG-6-30-15 observational data in 1994\footnote{In the following discussion, we assume that there is no narrow core
 component in the line profile, which is made of the line emission coming from the gas with small velocity, i.e.
 located far from the central black hole.}.  One can see excellent agreement between an observed profile and a theoretically calculated
 profile, especially in their red wing, for the viewing angle of $29.4^{\circ }$.  Although there are many previous results
 which agree with the data, this is the first calculation performed taking into account the fundamental physics which may operate
 the X-ray illumination in an accretion disk system, and we do not assume any phenomenological
 parameters such as the power-law index of the emissivity distribution.  This coincidence strongly supports our view of
 magnetic reconnection-heated corona.

Figure 5 shows the fits of our theoretical line profiles to more recent XMM-Newton observation of MCG-6-30-15 in 2001.
  Both line profiles are intrinsically the same, but made with different ways of subtraction of power-law continuum component.
  Roughly speaking, the left profile was obtaiend ignoring any contribution from the red component while the right profile
 was designed for the red tail to extend down to $\sim 3{\rm keV}$ 
 (see the caption under the figures for the detailed explanation of the line profiles).  As these fits show, our model
 can explain the recent observational line profile with high signal-to-noise ratio in the energy range of 4-7keV, though
 it fails to explain the red wing found below $\sim 4{\rm keV}$ (see the next section for the discussion).
       
With our model, we can calculate both the strength of continuum emission (LMO03) and that of iron line emission, so we can
 evaluate the equivalent width of the iron line emission in the context of our model.  Figure 5 shows the inclination dependence
 of the equivalent width of the iron K$\alpha$ line.  Especially, the equivalent width of the best-fit profile of MCG-6-30-15
 observation in 1994 is $\sim 70{\rm~eV}$, which is much lower than the value implied in the analyses of Tanaka et al. (1995),
 250-400 eV (i.e., we can fit the theoretical iron line profile with the observation, but fail to fit its absolute strength).
 However, this does not mean the failure of our model (see below).

\section{DISCUSSION AND CONCLUSION}
Using Monte Carlo simulation, we derive the X-ray irradiation from magnetic reconnection-heated corona onto the underlying
 disk, and find that the predicted iron line profile is consistent with past observations.  According to our calculation, the
 line emissivity on the accretion disk is approximately proportional to $r^{-5}$ where $r$ is
 the distance from the central black hole.

  As derived in LMS02, the fraction of accretion energy dissipated into the corona
 is almost unity.  In the standard model, the energy flux from the accretion disk is roughly proportional to $r^{-3}$, so one would
naively expect that the coronal illumination energy on the disk is also proportional to $r^{-3}$.  So how could such a steep
 emissivity profile be derived?

  The answer to this question is obvious from the behavior of Compton $y$-parameter, which is
 approximately proportional to $r^{-1}$.  The inward increase of $y$ can be understood in this way; the coronal temperature
 $T$ and density $n$ increase inward since in the inner region the coronal heating and chromospheric evaporation is more active
 than in the outer region.  Now $y=4kTn\sigma_T l/(m_e c^2)$, so $y$ also increases inward.  In an optically thin corona,
 the spectral index can be expressed as
$\alpha=(9/4+4/y)^{1/2}-3/2$.  Hence the spectrum of coronal radiation gets flatter as the distance from the
 black hole gets smaller.  As a result, the fraction of photons whose energy is high enough to drive iron fluorescence decreases
 outward.  Such an effect results in a steep line emissivity profile on the disk, and makes the
 red wing of an iron line profile rather prominent.

As seen in the recent observations, MCG-6-30-15 also showed a very broad iron line profile which had a significant red wing extending
 to $\sim 2{\rm keV}$ (for example, see the right panel of Figure 5).
  Wilms et al. (2001) concluded that, to account for such a broad line profile, the line emissivity profile should be as steep as
 a power-law $\propto r^{-\beta}$ with $\beta\sim 4-5$.  Some authors speculate that such a steep emissivity is the evidence
 of the extraction of rotation energy from the central black hole by Blandford-Znajek process (Blandford \& Znajek 1977).
  Our model can reproduce a emissivity profile which is roughly proportional to $r^{-5}$, though it is flattened inside $r\simeq 6r_g$.
  In order to reproduce such extended red wing in the iron line profile, which we gave up making a fit with our model perfectly, it may be
 necessary to assume the disk around
 a Kerr black hole (Dabrowski et al. 1997) or non-zero torque in the central region of the disk (Reynolds et al. 2004).  We can
 construct the corona model with disk models taking into account these assumptions (Novikov \& Thorne 1973; Agol \& Krolik 2000)
 and our model may have a great advantage in fitting the XMM-Newton data shown in Fig. 5 because of the steep emissivity law derived from it.  

We saw in \S 3 that the equivalent width of iron line emission evaluated in our model is much lower than that estimated from
 observational data.  The explanation for this discrepancy may be given by assuming the overabundance of iron (this possibility has already
 been mentioned
 in Tanaka et al. 1995).  Because of the lack of information about
 the iron abundance in MCG-6-30-15, we do
 not discuss this possibility further here. 

In this paper, we only show the fitting
 of the time-averaged line profile of MCG-6-30-15 with the time-stationary model.  In the observation, however, the iron line profiles from
 AGNs show significant time variability (Iwasawa et al. 1996, 1999).  Such time dependence is naturally expected in the framework
 of magnetic reconnection-heated corona, since magnetic energy dissipation is likely to occur sporadically, as is easily inferred
 by the analogy to solar flares.  Formation of blobs is also likely.  By taking into account rapidly moving
 blob-like coronae or magnetic flares as X-ray sources other than steady corona, we may explain the spectral variability
 of AGNs (Beloborodov 1999; Lu \& Yu 2001; Miniutti \& Fabian 2004).  In order to describe the generation and motion
 of such compact X-ray sources in the context of our model, it is necessary to perform time-dependent analyses of magnetic
 field generation and mass evaporation in the disk-corona system.
 Moreover, blob-like coronal structure may be favored when explaining
 observed large equivalent width of iron K$\alpha$ line, since such
 compact X-ray source would see the reflecting cold material (the
 disk) with relatively large solid angle, which makes the reflection
 fraction larger than in the case of a plain-parallel corona.  These topics are the beyond the scope of this paper,
 and left as future works.

\acknowledgements

We are grateful to B. F. Liu for providing part of the simulation code,
and to T. Tsuru, S. Nagataki, K. Ohsuga,
 K. Watarai, Y. Kato, A. Mizuta and R. Takahashi for their useful discussions and comments.
  This work was supported in part by the Grants-in Aid of the
Ministry of Education, Science, Sports, Technology, and Culture of
Japan (14079205,16340057, SM).

\clearpage

\vfill
\begin{figure}
\begin{center}
\resizebox{!}{10cm}{
\plotone{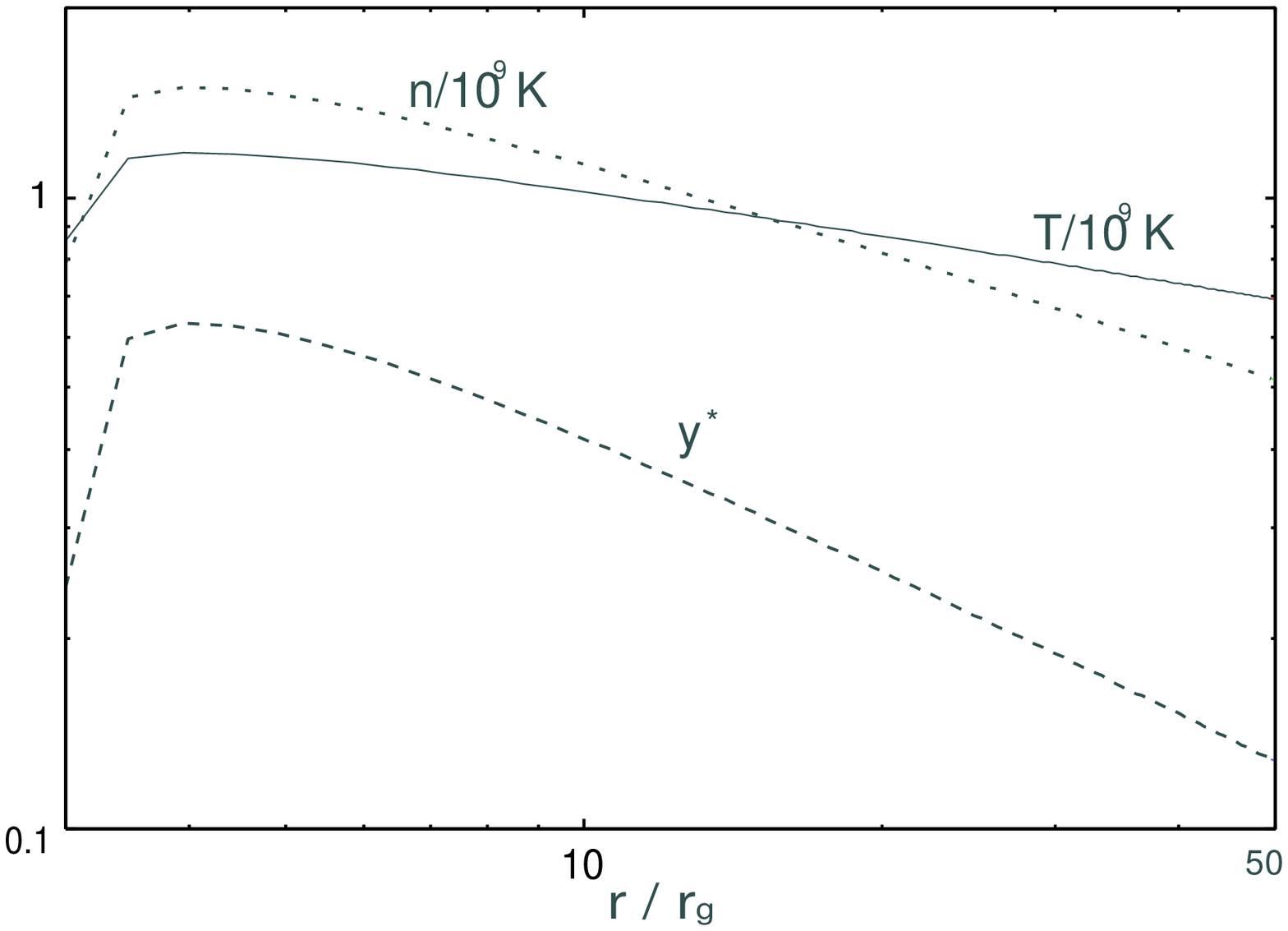}
}
\end{center}
\caption{Coronal quantities along distance for $M=10^8M_{\odot }$ and $\dot{M}=0.1\dot{M}_{\rm Edd}$. The coronal temperature is around
 $10^9{\rm K}$ and density around $10^9{\rm cm}^{-3}$.  $y^{*}=(4kT/mc^2)\lambda_{\tau}n\sigma_T l$ is the effective Compton
 $y$ parameter of the corona.}
\end{figure}
\vfill

\clearpage

\vfill
\begin{figure}
\begin{center}
\resizebox{!}{10cm}{
\plotone{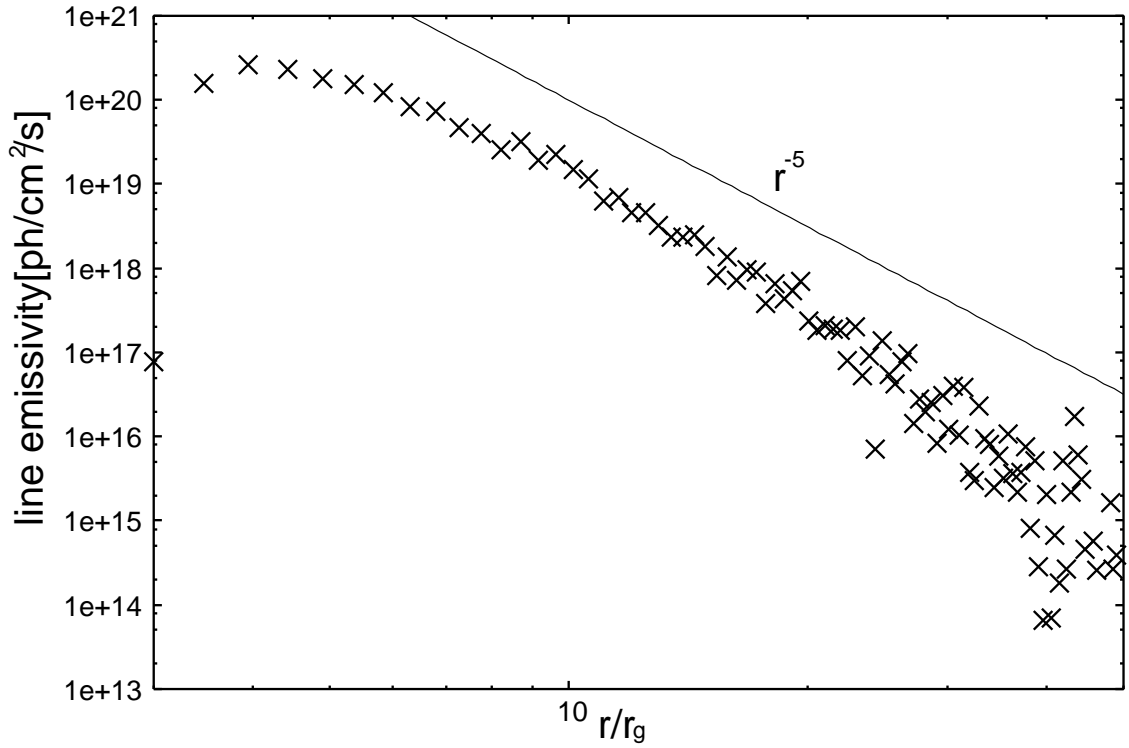}
}
\end{center}
\caption{The iron line emissivity profiles on the disk for $M=10^8M_{\odot }$ and $\dot{M}=0.1\dot{M}_{\rm Edd}$.
  The profile can be fitted with the power-law $\propto r^{-5}$.  This dependence is the same for different
 black hole mass or mass accretion rate.}
\end{figure}
\vfill

\clearpage

\vfill
\begin{figure}
\begin{center}
\resizebox{!}{8cm}{
\plotone{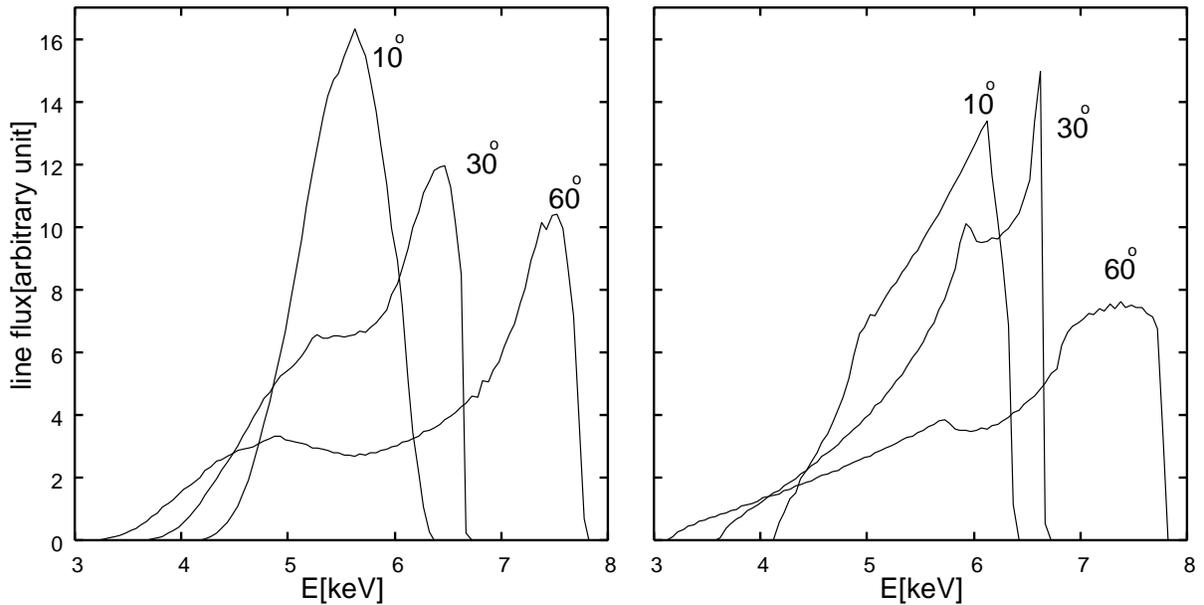}
}
\end{center}
\caption{Iron line profiles resulting from the accretion disk with magnetic reconnection-heated corona (left panel)
 and from simple power-law ($\propto r^{-3}$) emissivity distribution (right panel) are shown.  The black hole mass and the mass accretion
 rate are assumed to be $10^8M_{\odot }$ and $0.1\dot{M}_{\rm Edd}$, respectively, and the outer radius of the disk is assumed to be $50r_g$.
  Three inclinations are shown: $10^{\circ }$, $30^{\circ }$, and $60^{\circ }$.}

\end{figure}
\vfill

\clearpage

\vfill
\begin{figure}
\begin{center}
\resizebox{!}{10cm}{
\plotone{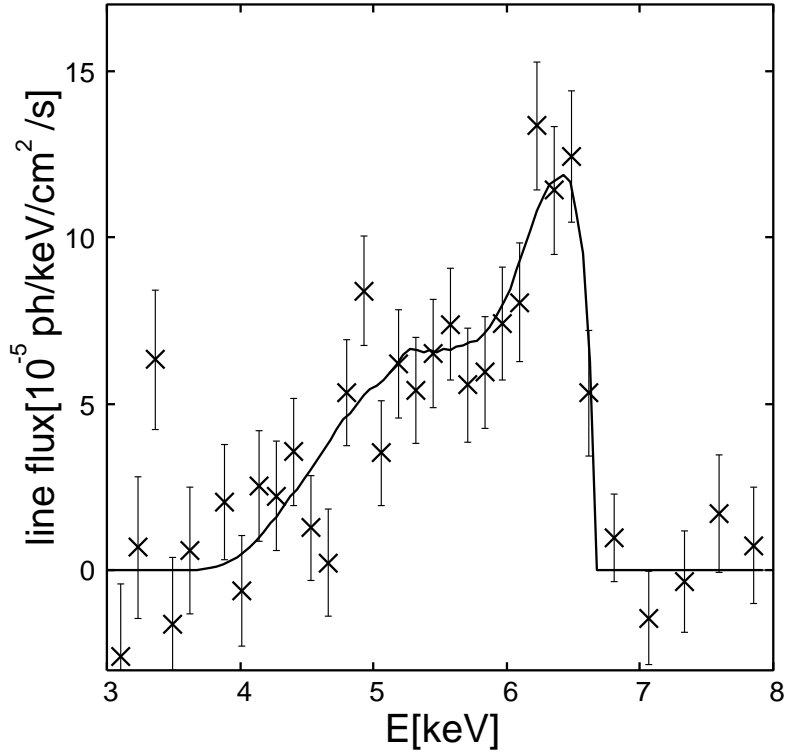}
}
\end{center}
\caption{Best-fit iron line profile for the time-averaged data of MCG-6-30-15.  This fit was made to unfolded data of ASCA observation in 1994,
 which is shown in Tanaka et al. (1995), using $\chi^2$ minimization.
  The viewing angle is $29.4^{\circ }$.}
\end{figure}
\vfill

\vfill
\begin{figure}
\begin{center}
\resizebox{!}{7cm}{
\plotone{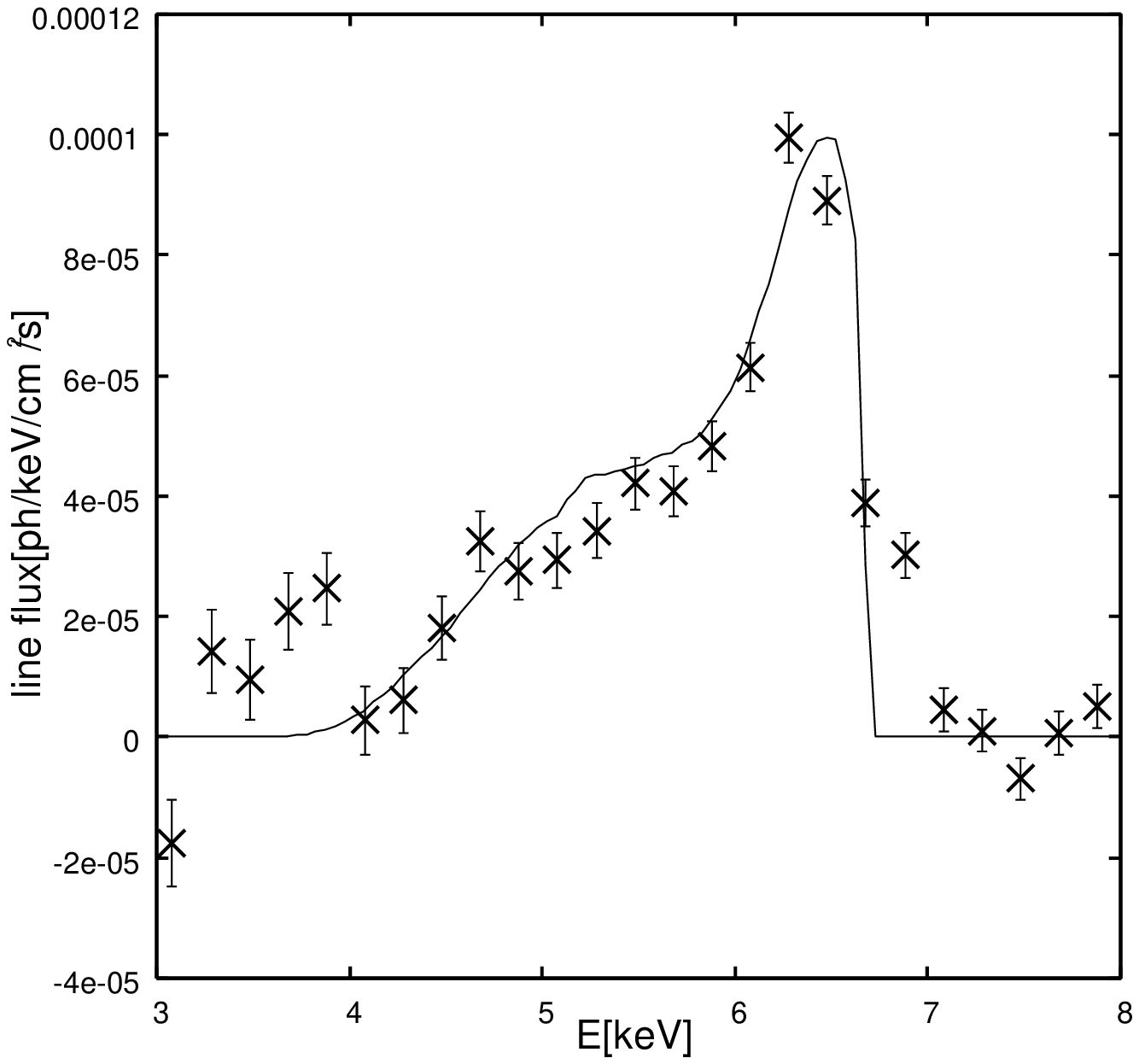}
\plotone{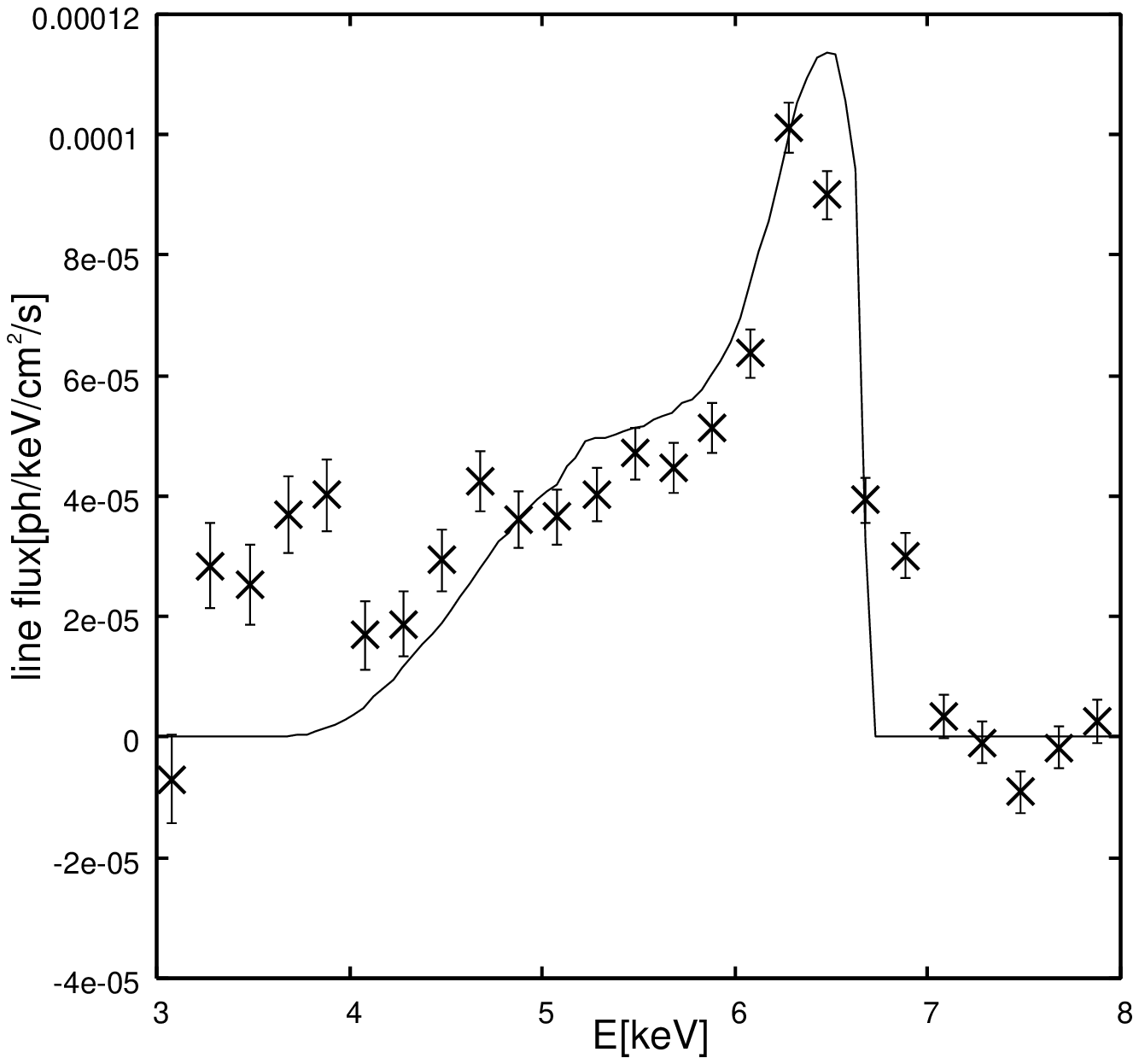}
}
\end{center}
\caption{Iron K$\alpha$ line profiles constructed from the long XMM-Newton observation of MCG-6-30-15 in 2001 (the total
 exposure time is 330ks, and the data were obtained from the EPIC pn camera) and the theoretical profiles which fit best to those are shown.
  The two profiles are excess emission above
 an absorbed power-law continuum, and the left profile is made excluding the 4.5-7.5keV, while the right profile is made
 excluding 3.2-7.5keV.  We make the $\chi^2$ minimization fits to our these observational line profiles with the viewing angle of $30.9^{\circ}$,
 though the normalization factors are different, using the data with the energy range of 4-7keV.
}
\end{figure}
\vfill

\vfill
\begin{figure}
\begin{center}
\resizebox{!}{10cm}{
\plotone{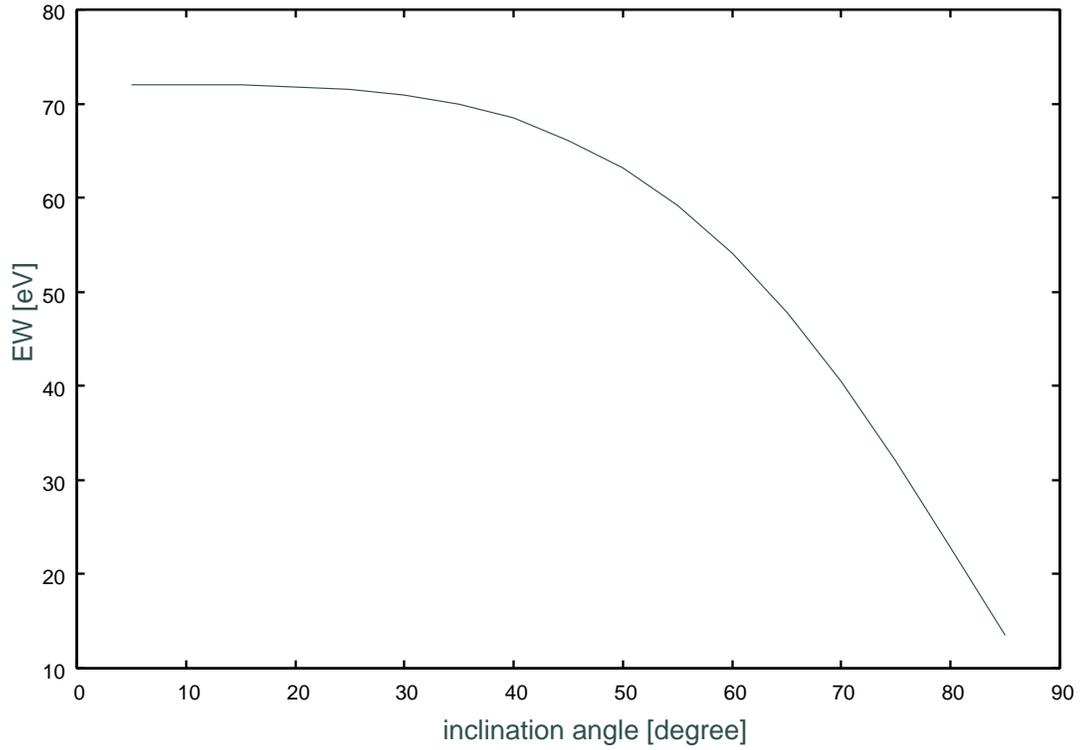}
}
\end{center}
\caption{The equivalent width (EW) of iron K$\alpha$ line emission
versus inclination angle, evaluated by our model with
$M=10^8M_{\odot}$ and $\dot{M}=0.1\dot{M}_{\rm Edd}$ (though the result is
not sensitive to the black hole mass and the mass accretion rate).}
\end{figure}
\vfill
      
\end{document}